\documentstyle[12pt,epsf]{article}
\begin{document}

\noindent
\begin{center}
\begin{large}
{\bf {Upstream Plasticity and Downstream Robustness} \\
{in Evolution of Molecular Networks.}}
\vspace{1.0cm}
\end{large}
\vspace{1.0cm}

\noindent {\bf Sergei Maslov$^{*)}$}\\
{\sl Department of Physics, Brookhaven National Laboratory, \\
Upton, New York 11973, USA}\\
Email: maslov@bnl.gov\\
\vspace{1.0cm}

{\bf Kim Sneppen}\\
{\sl Nordita, Blegdamsvej 17, 2100 Copenhagen {\O},
Denmark}\\
Email: Sneppen@nbi.dk\\
\vspace{1.0cm}

{\bf Kasper Astrup Eriksen}\\
{\sl Department of Theoretical Physics, Lund University,\\
S{\"o}lvegatan 14A, SE-223 62 Lund, Sweden.}\\
and\\
{\sl Department of Physics, Brookhaven National Laboratory, \\
Upton, New York 11973, USA}\\
Email: kasper@thep.lu.se.\\

\end{center}
\vspace{5.0cm}

\noindent
$^{*)}$Corresponding Author
\vfill\eject

\begin{Large}
\noindent
{\bf Abstract}
\end{Large}
\\\\

\noindent
{\bf Background:}\\
Gene duplication followed by functional
divergence of associated proteins is
a major force shaping molecular networks in living organisms.  Recent
availability of system-wide data for yeast S. Cerevisiae allow us
to access the effects of gene duplication on robustness and
plasticity of molecular networks.
\\\\

\noindent
{\bf Results:}\\
We demonstrate that the upstream transcriptional
regulation of duplicated genes diverges fast, losing on average 4\%
of their common transcription factors for every 1\% divergence of
their amino acid sequences. In contrast, the set of physical
interaction partners of their protein products changes much slower.  The
relative stability of downstream functions of duplicated genes, is further
corroborated by their ability to substitute for each other in
gene knockout experiments.
\\

\noindent
{\bf Conclusion:}\\
Apparently the upstream regulation of
genes evolves much more rapidly than the downstream functions of the
associated proteins. This is in accordance with a view where it is 
regulatory changes that mainly drives evolution. Any evolutionary 
model has eventually to account for this disparity and we have here 
quantified its size on a genome wide scale.
In this context a very important open question is to
what extent our results for duplicated genes within yeast (paralogs) 
carries over to homologeous proteins in different species (orthologs).
\\\\

\\\\

\newpage
\begin{Large}
{\bf Background}
\end{Large}
\\\\

Biological processes are rarely performed by single isolated
molecules. Instead, they typically involve a coordinated
activity of
many molecules
forming a neighborhood in biomolecular networks.
Changes in molecular networks are thus coupled to evolution of
new functions and functional relationships in the organism.
Gene duplication is an important source of raw material for the
evolutionary development of molecular networks in
a given species \cite{ohno1970}.
Immediately after a duplication event the pair of duplicated genes
is thought to be identical in both sequences and functional roles in the
cell. However, with time their properties including their position
within a network diverge. Here we quantify
this divergence in the yeast {\it Saccharomyces Cerevisiae}
using several recent system-wide data sets.
To this end we measure:
1) The similarity of positions of duplicated genes in the transcription
regulatory network \cite{lee2002} given by the number of
transcription regulators they have in common; 2) The similarity of
the set of binding partners \cite{uetz2000,ito2001} of their
protein products, and their ability to substitute for each other
in knock-out experiments \cite{giaever}. These measures reflect,
correspondingly, the upstream and downstream
properties of molecular networks around the duplicated genes.
\\\\

\begin{Large}
{\bf Results}
\end{Large}
\\\\

The first measure of divergence of duplicated
genes compares sets of their transcriptional
regulators. Such a set contains information about
different conditions under which the gene is expressed, and thus
reflects its functional roles in the cell.
To quantify the similarity of transcriptional regulation of a pair of genes
we introduce the concept of ``regulatory overlap'' $\Omega_{reg}$ given by
the number of transcription factors that bind to the
upstream regions of {\it both} these genes (see Fig. 1 for a
general illustration).
\begin{figure}[ht]
\epsfxsize=3in
\epsffile{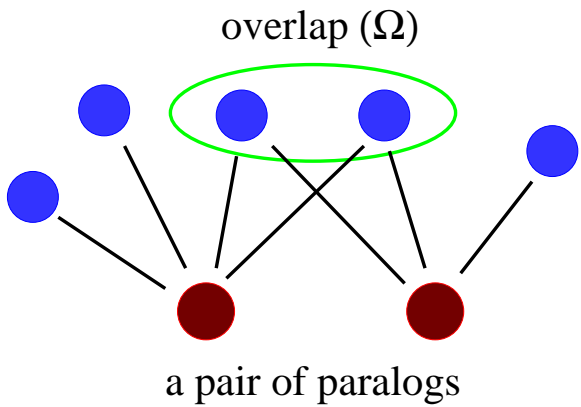}
\caption{Illustration of the concept of overlap in a molecular network.
For a pair of paralogs the overlap
$\Omega$ is defined as the number of common neighbors they have in the network.
In the case of transcription network the regulatory overlap $\Omega_{reg}$
counts transcription factors regulating both paralogs,
while for the physical interaction network
the interaction overlap $\Omega_{int}$ counts their common binding partners.
The pair of paralogs used in this illustration has the overlap $\Omega=2$
out of the total of 5 distinct neighbors of the pair. That
corresponds to a normalized overlap of $2/5=0.40$.
}
\end{figure}
The information about duplicated genes used in this study
was extracted from the list of all pairs of paralogous
(evolutionary related) proteins found in the yeast genome \cite{gilbert2002}, while the
system-wide data for its transcription regulatory network
was taken from the Ref. \cite{lee2002} (see Methods for more details).
Fig. 2A shows the distribution of the regulatory overlap for
different values of the percent identity (PID) of amino acid
sequences of paralogous proteins. 
\begin{figure}[tbh]
\epsfxsize=2in
\epsffile{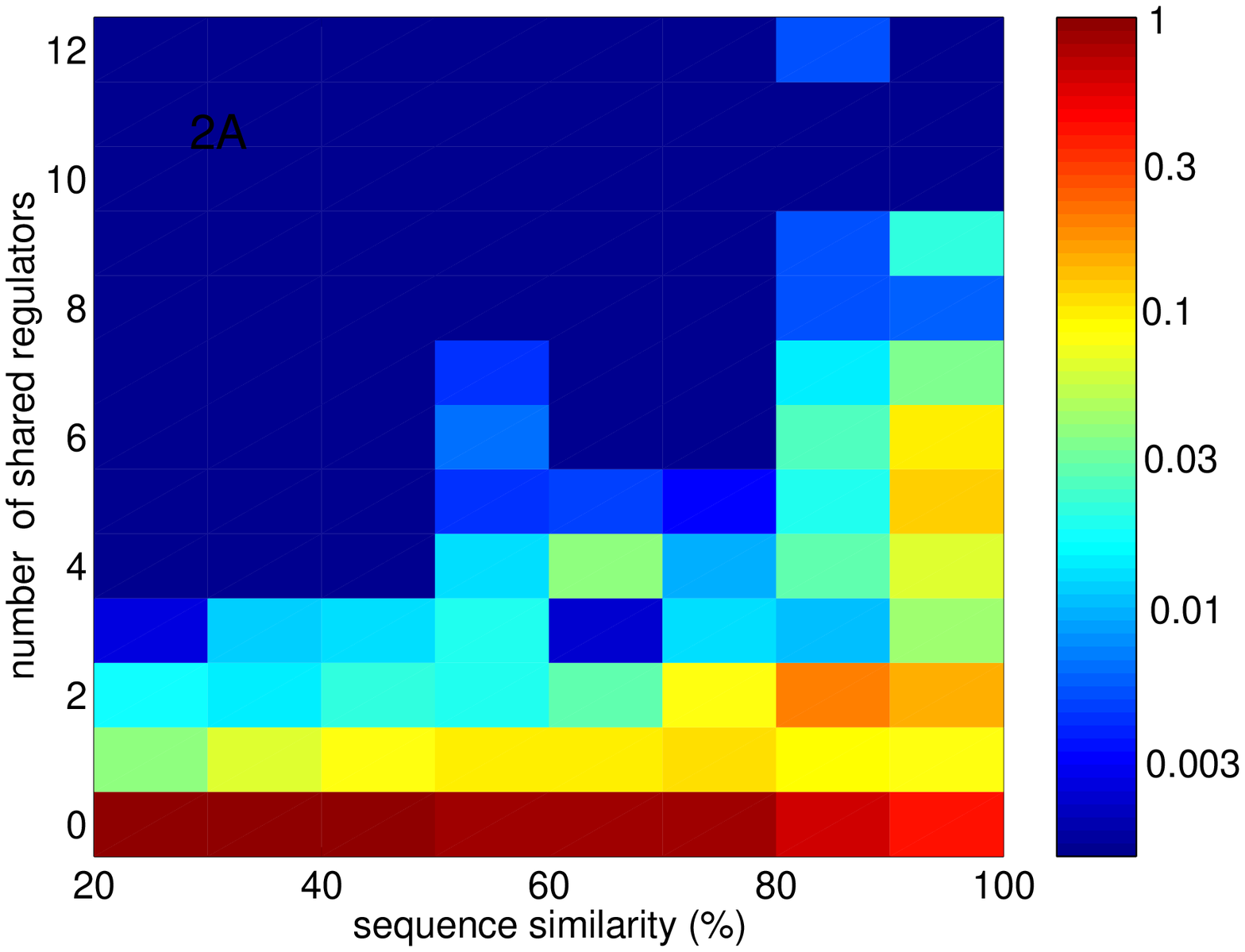}
\epsfxsize=2in
\epsffile{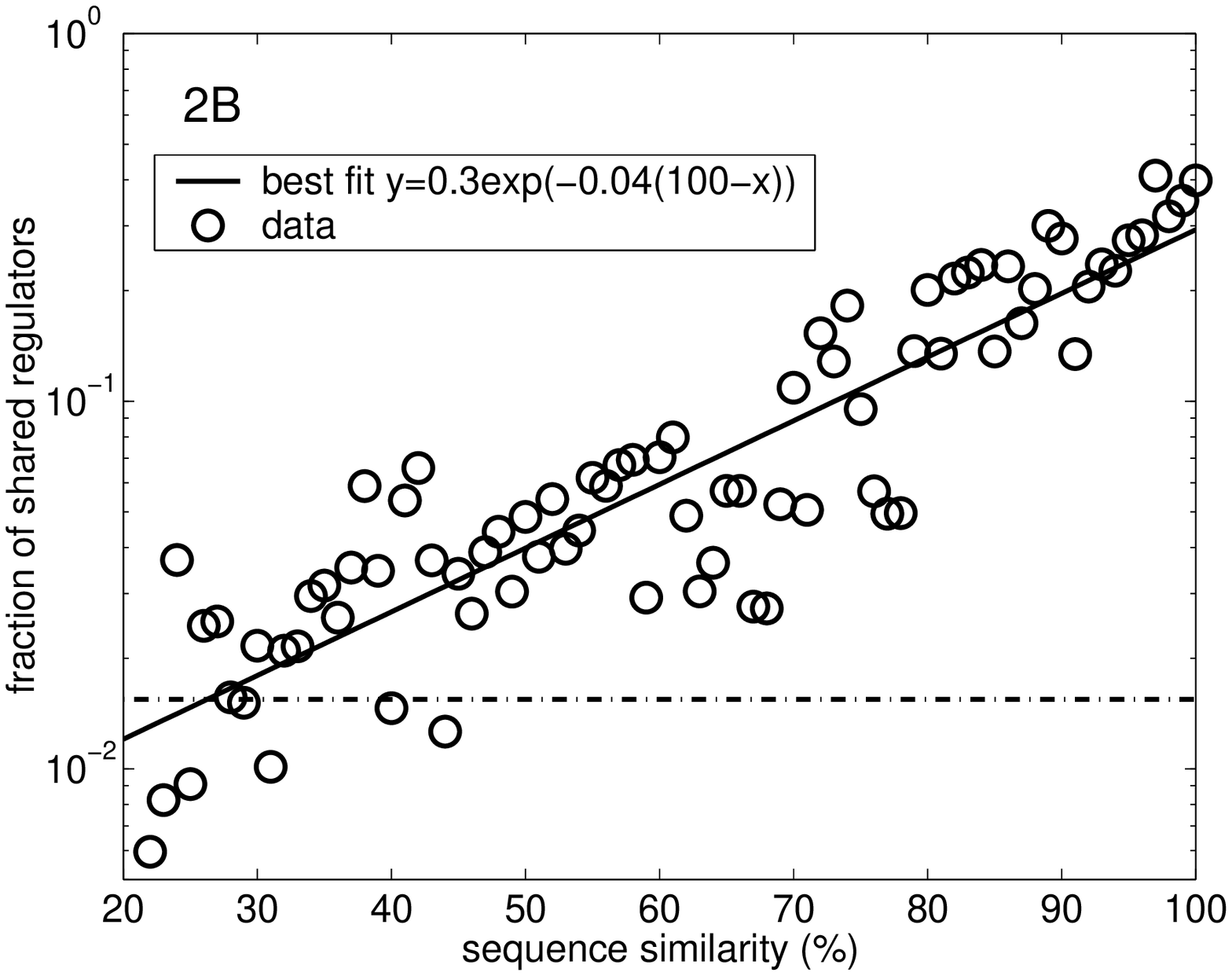}
\caption{
Divergence of the upstream regulation of duplicated genes.
A) The distribution of the number $\Omega_{reg}$ of
transcriptional regulators shared by a pair of paralogs,
as a function of the percent identity (PID) of their amino acid sequences.
Regulation data are taken from Ref. \cite{lee2002},
with the P-value threshold = 0.001.
B) The average regulatory overlap $\Omega_{reg}$ normalized
by the number of transcription regulators that regulate
either one or the other of the two genes as a function of the PID.
The solid line is a best fit to the exponential form
$\omega_0 \exp[-\gamma(100-PID)]$ with $\omega_0=0.3$, and $\gamma=0.04$.
The dashed horizontal line at $\omega_r=0.015$ is the 
normalized regulatory overlap of two random genes. 
}
\end{figure}
From this figure one can see
that the regulatory overlap decreases as a function of the PID.
While multiple overlaps dominate the distribution for PID $\geq$
80\%, at lower values of PID they disappear in favor of smaller
overlaps.

To quantify the average rate of loss of
the regulatory overlap $\Omega_{reg}$,
Fig 2B shows its
average value as a function of PID.
The regulatory overlap in this plot is normalized by the
ancestral connectivity of a gene, estimated as the total number of
distinct transcription factors that are involved in regulation
of at least one of the pair of proteins (see Fig 1).
One interesting feature of Fig. 2B is that even pairs
of proteins whose amino acid sequences are 100\% identical to each other
have only 30\% overlap in their upstream
regulation. The overlap becomes 60\% when measured
in units of the smaller among the two numbers of regulatory inputs
of a pair of paralogs.
The second feature of Fig. 2B is a gradual decline
of the average regulatory overlap over the whole range of
sequence similarities.
The data in Fig. 2B may be fitted with an exponential decay
with a rate corresponding to a 4\% chance of losing a given
common regulator of a pair for every 1\% decrease in their
amino acid sequence identity.
Thus already at PID$=$80\% half of the regulatory overlap
present at PID$=$100\% is lost.
The decline in the regulatory overlap presented
in Fig. 2A,B  is in accord with a recently published analysis
of similarity of microarray expression patterns \cite{gu2002}
of paralogs. In fact, due to a more direct
information about the gene regulation contained in the dataset of Ref.
\cite{lee2002}, our analysis extends the gradual decline to
much lower PID than could be detected \cite{gu2002}
from the microarray data. 
After we submitted this manuscript for the first time another study has
observed a rapid decline in the overlap of upstream motivs
\cite{TIG_papp_2003}. As this study 
was carried out as a function of Ks the long time behaviour 
can not be extracted.

We now consider the second measure of divergence, thus concentrating
on downstream functional properties of duplicated genes.
Such properties are in part reflected in their set of physical interaction partners
of their protein products.
The similarity of physical interaction neighborhoods of a pair of proteins
can be quantified as the ``interaction overlap'' $\Omega_{int}$ given by
the number of proteins that bind to both of them (See Fig. 1).
In our study we use the system-wide information about protein-protein
physical interactions obtained by combining two high throughput two-hybrid datasets
\cite{uetz2000,ito2001}.
Fig 3A shows the average value of the interaction overlap $\Omega_{int}$
between duplicated genes. 
Again it decreases with decreasing PID,
reflecting gradual loss/change of physical
interaction partners of proteins in the course of evolution.
\begin{figure}[tbh]
\epsfxsize=2in
\epsffile{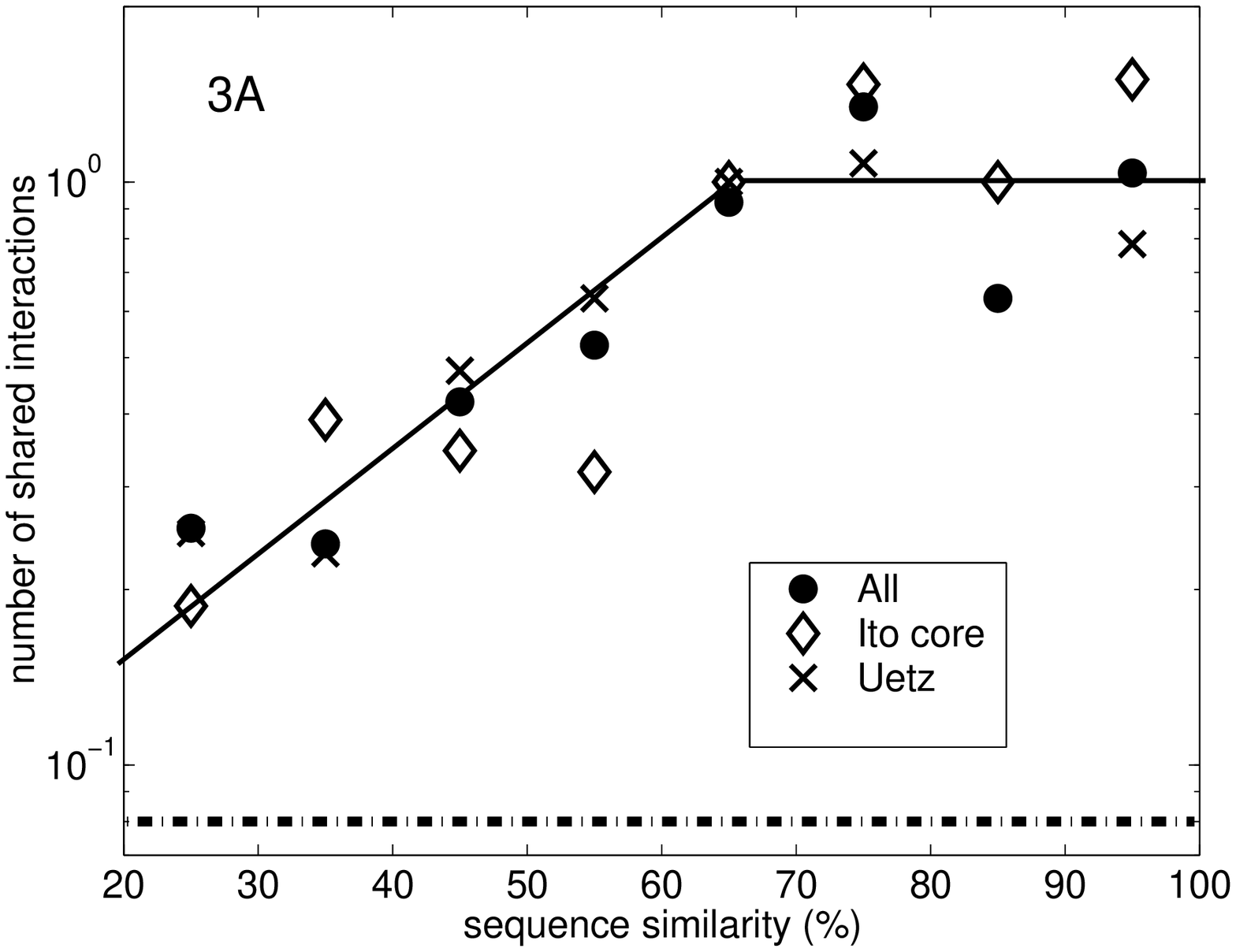}
\epsfxsize=2in
\epsffile{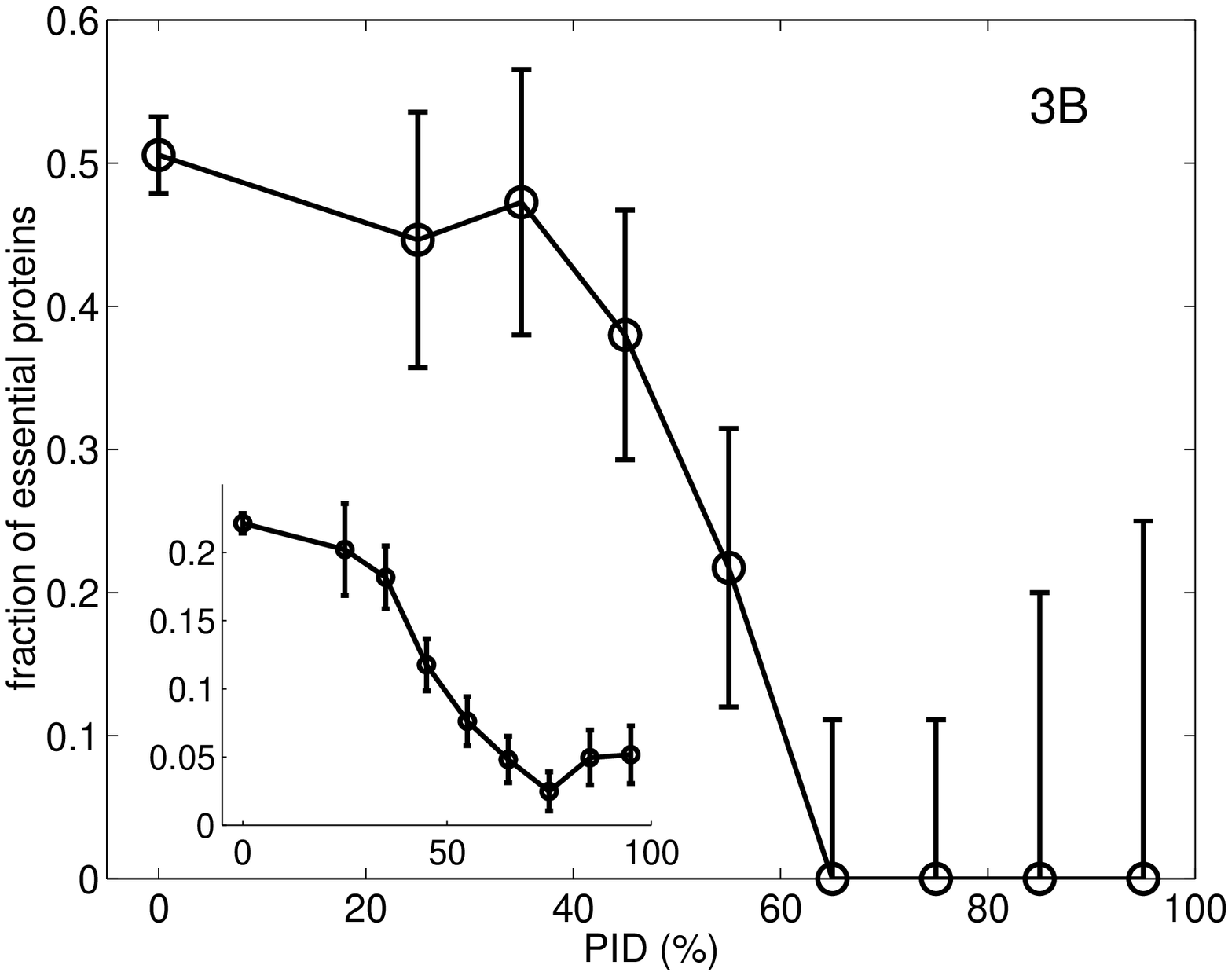}
\caption{Divergence of the downstream function of duplicated
genes. A) The average value of  the interaction overlap
$\Omega_{int}$ -- the number of protein interaction partners
shared by a pair of paralogs -- as a function of the similarity of
their amino acid sequences. The physical interaction data are
taken from the set of Uetz {\it et al.} \cite{uetz2000} (crosses),
core dataset of Ito {\it et al.} \cite{ito2001} (diamonds), and
the non-redundant combination of the two (filled circles). Note
the apparent plateau for PID's between 60\% and 100\% in both
datasets. Solid lines are guides for the eye, while the dashed
horizontal line at $8 \times 10^{-3}$ is the null-model
expectation value of the overlap. B) The fraction of essential
proteins among all proteins tested in Ref. \cite{giaever}, that
are also known to be localized in the yeast nucleus \cite{sgd}, is
plotted as a function of the PID to their most similar paralog in
the yeast genome. Proteins with no paralogs (singletons) are
binned at 0\% PID.
The insert (note the change of scale on the y-axis)
shows the fraction of essential proteins
among all (nuclear and non-nuclear) proteins, which is in agreement
with findings earlier reported in Ref. \cite{gu2003}.
}
\end{figure}
A similar analysis,
but as a function of $K_s$ --
the number of silent nucleotide substitutions per site --
was previously reported  by Wagner \cite{wagner2001}.
In agreement with that study, we find that gene duplicates
are more likely to share interaction partners
than one expects by pure chance alone (the horizontal line in
the Fig. 3A). Our set of yeast paralogs contains 366
pairs with both proteins present in the combined
set of
\protect{\cite{uetz2000,ito2001}}.
Out of these pairs 138
(38\%) share at least one interaction
partner.
We find that the decrease in average overlap becomes
systematic only for PID$<60$\%,
while above 60\% it remains
roughly constant in both Uetz \cite{uetz2000},
Ito \cite{ito2001}, and combined datasets (Fig. 3A).
Unlike in Fig. 2B, the overlap shown in Fig 3A is not normalized
by the total number of distinct interaction partners of
a paralogous pair.
The corresponding normalized plot confirms the above main conclusions.

An alternative way to quantify the extent of the divergence/redundancy
of downstream functions of a pair of duplicate genes is to examine the
viability of a null-mutant lacking one of them.
Gu, {\it et al.}  \cite{gu2003} recently analyzed the fraction of yeast
genes essential for the survival of the cell separately for singleton
genes that lack a duplicated partner (paralog) in the
yeast genome and the genes that retained at least one such partner.
It was found that the fraction of essential genes
is approximately 4 times higher among singleton genes
than among ones protected by a highly similar paralog.
It was also demonstrated that the protective role of a paralog
persists down to rather low levels of the
amino acid sequence similarity (PID).
In Fig. 3B we confirm these findings using a different set of
lethal and viable null-mutants of Ref. \cite{giaever} as well as
demonstrate that the magnitude of the effect is the strongest
among nuclear proteins, where the largest fraction of essential
proteins resides. Notice that the lethality (especially that of
nuclear proteins) shows a dramatic change at PID around 60\%,
indicating that paralogs with a higher level of similarity can
typically substitute for each other.
\\\\

\begin{Large}
{\bf Discussion}
\end{Large}
\\\\

Having presented different measures of the upstream and the downstream
divergence of duplicated genes we are now in a position to
discuss them in a wider context.
Comparing Fig. 2B to Fig. 3A,B one concludes
that changes in the upstream regulation of duplicated genes
happen more readily than changes in their downstream function.
The overlap in the set of binding partners (Fig. 3A) and the
ability of duplicates to substitute for each other
(Fig. 3B) remain virtually constant down to PID of 60\%,
while the average regulatory overlap at this PID has dropped to
about 20\% of its maximum (Fig. 2B).
Thus our results indicate that duplicated genes
would still have the ability to
partially substitute for downstream functions of each other
at the time when the repertoire of their regulatory
connections has substantially changed.
Such genes would be less constrained in evolving new functions
\cite{kondrashov2002},  and thus would contribute to a greater evolutionary
plasticity of the network.
\\\\

\begin{Large}
{\bf Conclusions}
\end{Large}
\\\\

The evolution of a biological species organism modifies it on
multiple levels ranging from sequences of individual molecules, to
their coordinated activity in the cell (molecular networks), all
the way up to the phenotype of the organism itself. While its
manifestations both on the level of sequences and phenotypes are
well documented, the data needed to quantify the evolutionary
changes on the level of molecular networks have appeared only very
recently. System-wide studies such as high-throughput two hybrid
assays of protein-protein interactions \cite{uetz2000,ito2001},
the large scale study of transcriptional regulation
\cite{lee2002}, and the whole genome assay of the viability of
null-mutants in yeast \cite{giaever} have allowed us to go beyond
describing particular cases of evolution of molecular networks and
look at its large scale dynamics.

Our results show that genetic regulations of duplicated proteins
in yeast change faster than both their amino acid sequences and
their protein interactions partners. It is tempting to extend this
observation to pairs of homologous proteins in different species
(orthologs) that diverged from each other as a result of a
speciation (as opposed to gene duplication) event. This would help
to explain how species with very similar gene contents can evolve
novel properties on a relatively short timescale. However, such an
inter-species comparison of molecular networks has to wait for a
completion of large-scale studies of closely related model
organisms.
\\\\

\begin{Large}
{\bf Methods:}
\end{Large}
\\\\

As a source of information about duplicated genes we use the set
\cite{gilbert2002} consisting of 4443 pairs of paralogous yeast proteins.
This set was obtained by blasting all yeast proteins against each other
with a conservative E-value cutoff of $10^{-30}$.
We curated this dataset by removing 72 known
transposable elements, http://genome-www.stanford.edu/Saccharomyces,
and their
homologs (108 proteins all together).
That left us with  2739 pairs of paralogous yeast proteins formed by
1891 proteins (about 30\% of the genome) with at least one homolog
in the yeast genome.
Pairs of paralogous proteins in this set are characterized by a fairly
broad distribution of the percent identity (PID) of their
amino acid sequences in the interval from 20\% to
100\%,
as shown in Fig. 4.
\begin{figure}[tbh]
\epsfxsize=4in
\epsffile{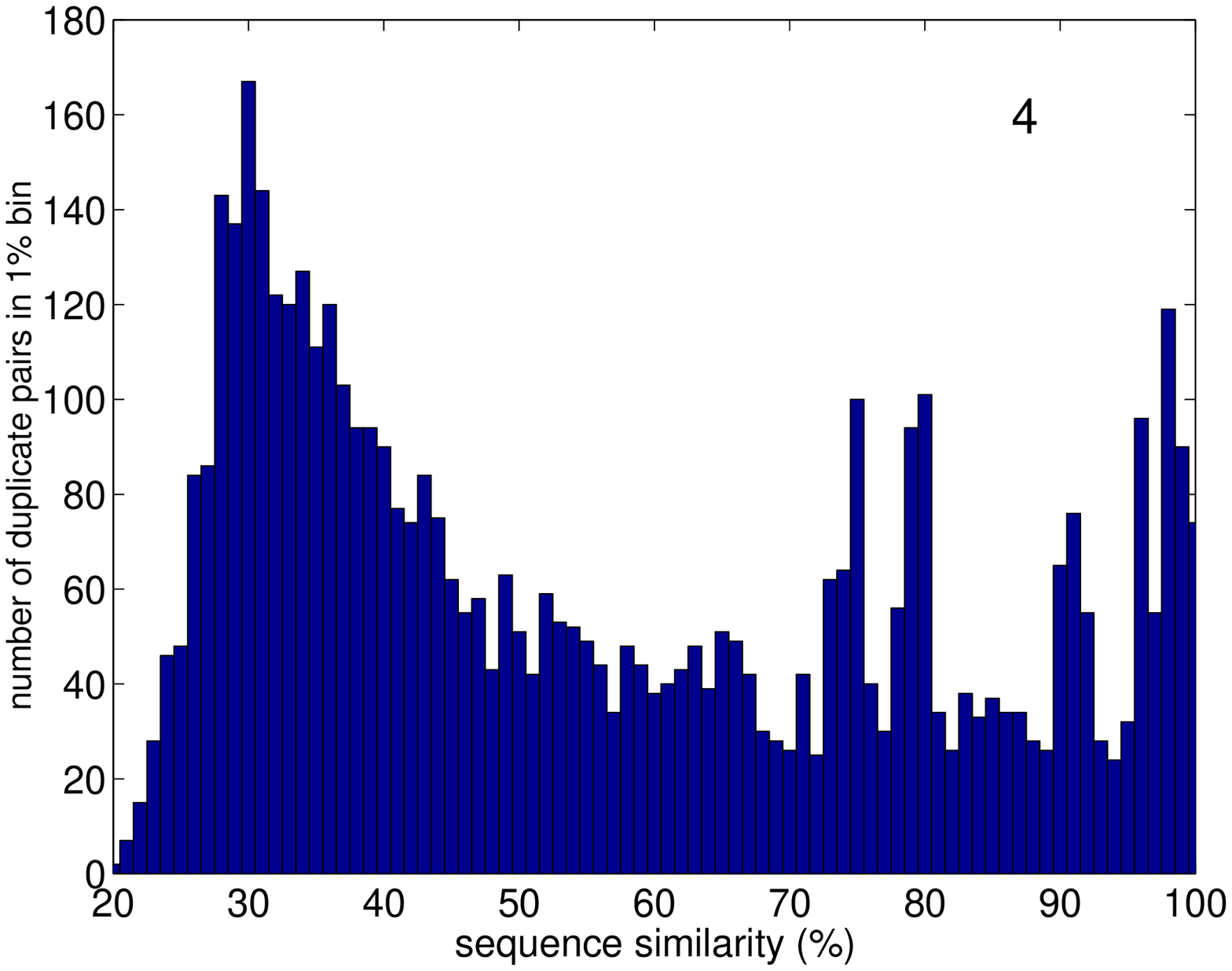}
\caption{ The histogram of the PID of the 2739 pairs of paralogous
proteins \protect{\cite{gilbert2002}} used in our study.}
\nonumber
\end{figure}
It is worthwhile to note that while
the number of silent substitutions $K_s$ per site in a pair of
duplicated genes is commonly used as a proxy of the time elapsed
since the duplication event \cite{ohno1970}, the PID (or the number
of non-silent substitutions per site $K_a$=1-PID/100)
is rather a very crude estimate of the extent of their functional
similarity.

The system-wide
data describing the transcription regulatory network of yeast
was taken from the Ref. \cite{lee2002}, which reports the
in-vivo study of binding between 106 transcription
factors to the upstream regulatory regions of genes encoding all
6270 of yeast proteins. Since the number of transcriptional regulators
in this dataset is quite large, the probability that by pure chance
the same transcription factor would be incorrectly detected among
upstream regulators {\it both} duplicated genes is relatively small.
Thus the contribution of false positives
to the regulatory overlap $\Omega_{reg}$ is insignificant.
On the other hand, false positives significantly affect the average
number of regulatory inputs of each individual proteins.
Thus the presence of a considerable
fraction of false positives would manifest
itself in sensitivity of the {\it normalized}
regulatory overlap with respect
to the P-value cutoff of the data.
We failed to find any such dependence when investigating
regulatory overlap for different P-values (data now shown):
While the average number of regulations per gene decreases six-fold
(from 2 to 0.33) when the P-value cutoff is lowered from $10^{-2}$ to $10^{-4}$,
both the initial drop and the rate of the exponential decay fit in Fig. 2B remains virtually
unchanged. This suggests that false positives are not a significant
part of the experimental dataset of Ref. \cite{lee2002} at least up to $10^{-2}$,
and validates the robust nature of the parameters extracted from the Fig. 2B.

As a source of information about binding partners of
yeast proteins we combine two high throughput two-hybrid datasets:
the core Ito {\it et al.}  set \cite{ito2001}
(806 interactions among 797 proteins)
and the extended Uetz {\it et al.} dataset \cite{uetz2000} downloaded from
the website of this group (1446 interactions among 1340 proteins)
The resulting dataset consists of a total of 1734 proteins
joined by 2111 non-redundant interactions.
Using this combined dataset we found that even 100\% identical
proteins share on average only 30\% of their binding partners. However, unlike in
the case of the upstream regulation, the
set of interaction partners of a protein in a two-hybrid
experiment is essentially determined by its amino acid sequence. We attribute the
30\% overlap in the set of binding partners of identical proteins
to false positives/negatives inevitably present
in high-throughput two-hybrid experiments. The presence of
false positives/negatives also manifests itself in
the fact that two independent system-wide
two-hybrid experiments \cite{uetz2000}, \cite{ito2001}
have only 141 interacting pairs in common.

The system-wide data on viability of null-mutants
used in our study was obtained from Ref.
\cite{giaever} in which 1103 essential (non-viable null-mutants)
and 4678 non-essential (viable null-mutants) yeast proteins were
reported. Actual lists of viable and non-viable
null-mutants as discovered in Ref. \cite{giaever}
were downloaded from the Saccharomyces Genome Database
http://genome-www.stanford.edu/Saccharomyces.
\\\\

\begin{Large}
{\bf Authors Contributions}
\end{Large}
\\\\

All authors contributed to both the ideas and
writing the manuscript in close collaboration.
All authors read and approved the manuscript.
\\\\

\begin{Large}
{\bf Acknowledgment}
\end{Large}
\\\\

Work at Brookhaven National Laboratory was carried
out under Contract No. DE-AC02-98CH10886, Division
of Material Science, U.S. Department of Energy. Two of
us (K.E and K.S.) thank the Institute for Strongly
Correlated and Complex Systems at Brookhaven National
Laboratory for financial support during visits when part
of this work was completed. S.M. and K.S. acknowledge the
support of the NSF grant PHY99-07949 (work at the KITP,
University of California at Santa Barbara). We thank
John Little for critically reviewing the manuscript.

\vfill\eject

\end{document}